\documentclass[aps,prl, amsmath, showpacs, preprintnumbers,superscriptaddress, twocolumn,sort&compress,floatfix, amssymb]{revtex4}
\usepackage{graphicx}
\usepackage{rotating}
\usepackage{dcolumn}
\usepackage{bm}
\usepackage{color}
\usepackage{mathptmx, textcomp}
\usepackage[latin1]{inputenc}
\usepackage{braket}

\begin{document}

\author{M. J. Mark}
\author{E. Haller}
\author{K. Lauber}
\author{J. G. Danzl}\affiliation{Institut f\"ur Experimentalphysik and Zentrum f\"ur Quantenphysik, Universit\"at Innsbruck, 6020 Innsbruck, Austria}
\author{A. J. Daley}\affiliation{Department of Physics and Astronomy, University of Pittsburgh, Pittsburgh, PA 15260, USA}
\author{H.-C. N\"agerl}\affiliation{Institut f\"ur Experimentalphysik and Zentrum f\"ur Quantenphysik, Universit\"at Innsbruck, 6020 Innsbruck, Austria}

\title{Precision Measurements on a Tunable Mott Insulator of Ultracold Atoms}

\date{\today}


\begin{abstract}
We perform precision measurements on a Mott-insulator quantum state of ultracold atoms with tunable interactions. We probe the dependence of the superfluid-to-Mott-insulator transition on the interaction strength and explore the limits of the standard Bose-Hubbard model description. By tuning the on-site interaction energies to values comparable to the interband separation, we are able to quantitatively measure number-dependent shifts in the excitation spectrum caused by effective multi-body interactions.
\end{abstract}

\maketitle

The observation of the superfluid-to-Mott-insulator transition in the context of ultracold atoms \cite{Jaksch1998,Greiner2002} has triggered numerous activities both in theory and in experimental physics \cite{Bloch2008}. It has become clear that ultracold gaseous systems confined to optical lattice potentials are capable of serving as bottom-up models for condensed matter phenomena \cite{Jaksch2005,Lewenstein2007,Bloch2008}. In addition, in view of unprecedented control and read-out capabilities \cite{Bakr2010,Sherson2010,Weitenberg2011}, there is justified hope that ultracold atomic and molecular systems will allow the implementation of quantum simulation schemes \cite{Jane2003,Buluta2009}. While there is tremendous progress for fermionic systems confined to optical lattices \cite{Jordens2008,Schneider2008}, most experiments have so far addressed the case of bosonic quantum gases and in particular the quantum phase transition from a superfluid to an insulating Mott state \cite{Greiner2002,Stoeferle2004,Campbell2006,Gemelke2009,Clement2009,Fukuhara2009,Haller2010,Bakr2010,Sherson2010}. In this case, as long as the interaction can be treated as a weak perturbation, the system is described by the Bose-Hubbard (BH) model \cite{Fisher1989}. One of the merits of ultracold atomic systems is the fact that all parameters of the BH model can be derived in a microscopic way \cite{Jaksch1998}. Nevertheless, recent theoretical \cite{Luhmann2002,Johnson2009,Buechler2010} and experimental \cite{Will2010} investigations have shown that already for comparatively weak interactions corrections to the standard BH model are needed.

In this Letter we use our capability to tune interaction energies to values comparable to the interband spacing and thereby leave the range of validity for the approximations of the standard BH model description - specifically, the restriction that only the lowest Bloch band in the lattice is occupied, and the treatment of interactions via the zero-range pseudopotential applied in the Born approximation. Using a Bose-Einstein condensate (BEC) of Cs atoms loaded into a 3D optical lattice potential we first investigate the superfluid-to-Mott-insulator transition and its dependence on the interaction strength. We precisely determine the on-site interaction energies including effective multi-body interaction shifts, demonstrating the breakdown of the standard approximations. Our results show good agreement with treatments beyond the BH model incorporating both higher bands and regularization of the pseudopotential.

The standard BH model introduces two parameters to describe the dynamics of ultracold atoms in an optical lattice: the rate $J/\hbar$, which describes tunneling between neighboring lattice sites, and the energy $U$, which quantifies the interaction of atoms at the same lattice site. In the presence of weak external harmonic confinement the Hamiltonian reads
\begin{equation}\label{HamiltonBoson2}
\begin{split}
\widehat{H}=-J\sum\limits_{<i,j>}\widehat{a}_i^\dagger\widehat{a}_j+\sum\limits_i\frac{U}{2}\widehat{n}_i\left(\widehat{n}_i-1\right)+\sum\limits_i \epsilon_i \widehat{n}_i,
\end{split}
\end{equation}
where $\widehat{a}_i^\dagger$ ($\widehat{a}_i$) are the bosonic creation (annihilation) operators at the $i$-th lattice site, $\widehat{n}_i\,{=}\,\widehat{a}_i^\dagger\widehat{a}_i$ is the number operator, and $\epsilon_i$ denotes the on-site energy shift due to an external confinement. For small $U/J$ the ground state at zero temperature is a superfluid (SF), whereas for large $U/J$ and commensurate filling on-site interactions inhibit tunneling and the ground state is the Mott insulator (MI) of exponentially localized atoms. These limits are connected by a quantum phase transition. The transition point for a homogeneous system can be calculated in a mean-field approach, giving $(U/J)_{\rm c}\,{=}\,34.8$ in a 3D cubic lattice \cite{Fisher1989}, close to the quantum Monte-Carlo result $(U/J)_c\approx 29.3$ \cite{Sansone2007}.

In the standard BH model, $U$ and $J$ are usually calculated from lowest-band Wannier functions \cite{Jaksch1998}. Two-body interactions can be described via a regularized zero-range pseudopotential, as the system is dilute, and relative momenta between atoms are small compared with scales determined by the range of the interaction potential \cite{Tiesinga2000,Castin2001}. For small values of the s-wave scattering length $a_{\rm S}$, this pseudopotential is then replaced by a $\delta$-function in the integrals that determine $U$ (giving the Born approximation result). Under these approximations, the standard BH model was successfully used to describe a range of experiments with ultracold atoms in optical lattices \cite{Greiner2002,Bloch2008}. However, for sufficiently strong interactions the approximations break down. In a simple picture, two or more particles localized at a single lattice site with strong interactions tend to avoid each other and the on-site wave function increases in width to minimize the energy, resulting in coupling to higher Bloch bands. This admixture of higher bands results in number-dependent shifts for the on-site interaction energy, and the standard BH model may be modified to reproduce the new bound state energies \cite{Buechler2010} by replacing $U$ by a number-dependent term $U(n_i)$ \cite{Luhmann2002,Johnson2009,Will2010,Tunneling}. However, care must be taken, as the replacement of the pseudo-potential with a $\delta$-function is in general not valid when including higher bands, and instead it is necessary to use the full regularized zero-range potential \cite{Busch1998}. Note that small modifications for $U$ as a function of $n_i$ are already visible for weak interactions but reasonably deep lattices \cite{Campbell2006,Will2010}.

\begin{figure}
\includegraphics[width=8.5cm]{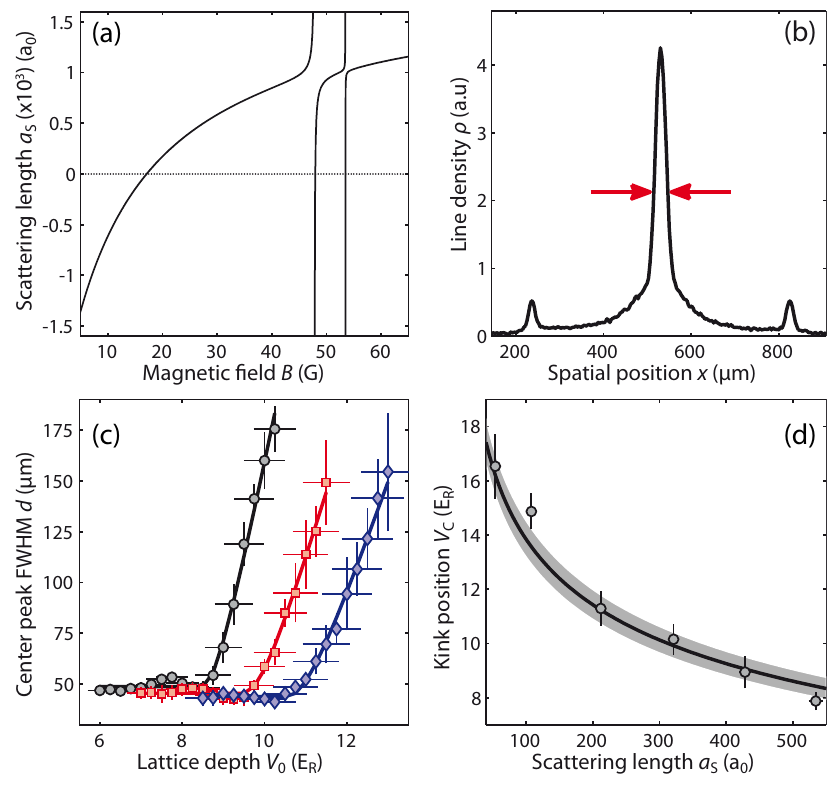}
\caption{(color online) a) Scattering length $a_{\rm S}$ for Cs atoms in the lowest hyperfine ground state as a function of the magnetic field $B$ as calculated in Ref.~\cite{Lange2009,Julienne2011}. b) Example of an integrated density profile of the BEC for $50\,$ms TOF. The arrows indicate the FWHM $d$. c) Center peak FWHM $d$ as a function of lattice depth $V_0$ for $427\,a_0$ (circles), $320\,a_0$ (squares), and $212\,a_0$ (diamonds). The solid lines are fits \cite{fit1} from which the critical lattice depth $V_{\rm C}$ is determined. d) Critical depth $V_{\rm C}$ as a function of $a_{\rm S}$. The solid line corresponds to the transition points for the SF-to-MI transition given by $(U/J)_{\rm c}\,{=}\,34.8$ \cite{Fisher1989}, the shaded area indicates our uncertainty for $V_0$. \label{Fig1}}
\end{figure}

The starting point for our experiment is a BEC without detectable non-condensed fraction of typically $1.0\times10^{5}$ Cs atoms in the energetically lowest hyperfine ground state confined by a crossed dipole trap. Atom cooling and trapping follow the procedures described in Ref.~\cite{Weber2003,Kraemer2004}. The cubic lattice is generated by three retro-reflected laser beams at a wavelength of $\lambda\,{=}\,1064.5\,$nm. With the given laser power the maximum lattice depth $V_0$ is $30\,E_{\rm R}$, where $E_{\rm R}\,{=}\,h^2/(2m\lambda^2)$ is the atomic recoil energy with the mass $m$ of the Cs atom. The strength of interactions can be widely tuned as $a_{\rm S}$ depends strongly on magnetic field $B$ \cite{Lange2009,Julienne2011} due to multiple Feshbach resonances as illustrated in Fig.\,\ref{Fig1}(a).

We first probe the transition from the SF to the MI state as we vary the strength of interactions, using the standard interference-contrast technique \cite{Greiner2002}. We gently load the BEC within $400$\,ms into the lattice, hold the atoms for $10\,$ms, and then instantly switch off both the lattice and the external trap to determine the momentum distribution in a $50\,$ms time-of-flight (TOF). We determine the FWHM $d$ of the central peak of the resulting interference pattern \cite{Wessel2004} as shown in Fig.\,\ref{Fig1}(b). As expected, $d$ shows a strong dependence on $V_0$ (see Fig.\,\ref{Fig1}(c)). We relate the onset of the MI phase to the abrupt kink in the data, corresponding to a critical depth $V_{\rm C}$ \cite{fit1}. Fig.\,\ref{Fig1}(d) shows $V_{\rm C}$ as a function of $a_{\rm S}$. We find remarkable agreement with the values for the calculated MI-transition points \cite{Fisher1989} for the case of a homogeneous system with integer density of one atom per lattice site. We note that in general our data on the set of transition points exhibits some dependence on the initial density. Here, the density is chosen such that for a given interaction strength the mean atom number per lattice site in the central region of the trap is near unity at the transition point.

\begin{figure}
\includegraphics[width=8.5cm]{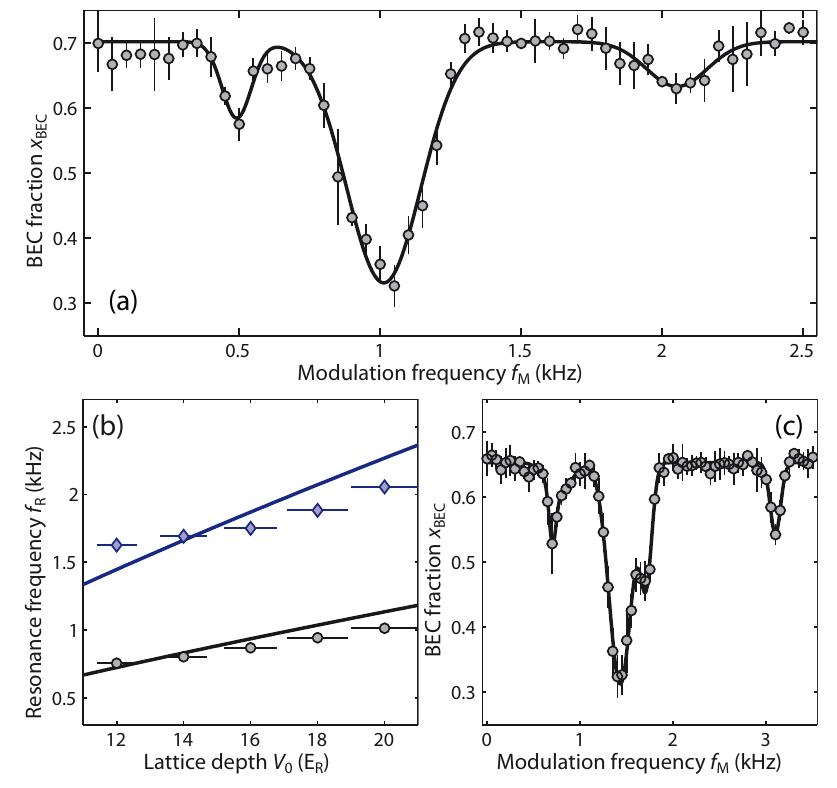}
\caption{(color online) a) Excitation spectrum in the MI phase at $V_0\,{=}\,20\,E_{\rm R}$ and $a_{\rm S}\,{=}\,212\,a_0$. The solid line is a triple gaussian fit. We take the resonance position as the center of the gaussian peak. b) Frequency of the $U$ resonance (circles) and the $2U$ resonance (diamonds) for various values of $V_0$ for $a_{\rm S}\,{=}\,212\,a_0$. The solid lines are the calculated frequencies corresponding to $U(1)$ and $2\,U(1)$. At this value for $a_{\rm S}$ the transition point occurs at $\approx\!12\,E_{\rm R}$. c) Excitation spectrum in the MI regime at $20\,E_{\rm R}$ and $a_{\rm S}\,{=}\,320\,a_0$. The solid line is a five-peak gaussian fit. \label{Fig2}}
\end{figure}

A second signature for the MI phase is the opening up of a gap and hence the appearance of distinct resonances in the excitation spectrum \cite{Greiner2002,Stoeferle2004,Clark2006,Kollath2006}. Here, we will find that the experiment deviates significantly from the results of the standard BH model and the associated approximations. Fig.\,\ref{Fig2}(a) shows a typical spectrum when the system is deeply in the MI phase. We plot the BEC fraction \cite{Naraschewski1998} after $150\,$ms of amplitude modulation (AM) at $20\,$\% of $V_0$ and back-transfer into the initial dipole trap as a function of the AM frequency $f_\text{M}$. The BEC fraction is a sensitive indicator for temperature changes and hence for the amount of energy deposited into the system. The spectrum in Fig.\,\ref{Fig2}(a), taken for comparatively weak interactions and low atom density, shows three characteristic peaks around the energies $U/2$, $U$ and $2\,U$. The $U/2$ peak relates to a two-phonon transition, while the $2\,U$ peak corresponds to an excitation at the edge between the singly and doubly occupied shells \cite{Clark2006,Kollath2006}. Interestingly, when we take spectra like the one in Fig.\,\ref{Fig2}(a), the peaks, in particular the one at $U$, are typically not well fit by symmetric gaussian functions. The reason for this will become evident below. Fig.\,\ref{Fig2} (b) plots the positions of the $U$ and $2\,U$ resonances as a function of $V_0$ for $a_\text{S}\,{=}\,212\,a_0$ and compares them to the results of the standard BH model. In general, the agreement is not satisfactory. For comparatively deep lattices (above $18\,E_{\rm R}$) we measure a significant downshift for both the $U$ and the $2\,U$ resonance. Near the transition point (here at $V_0\,{=}\,12\,E_{\rm R}$) the $2\,U$ resonance is clearly upshifted. The latter can be understood when taking the spatially separated coexistence of the SF and MI phase into account, as for the SF part the maximal excitation probability lies energetically above the expected value for $U$ and hence also for $2\,U$ \cite{Clark2006,Kollath2006}. In the following, we will find an explanation for the downshift.

\begin{figure}
\includegraphics[width=8.5cm]{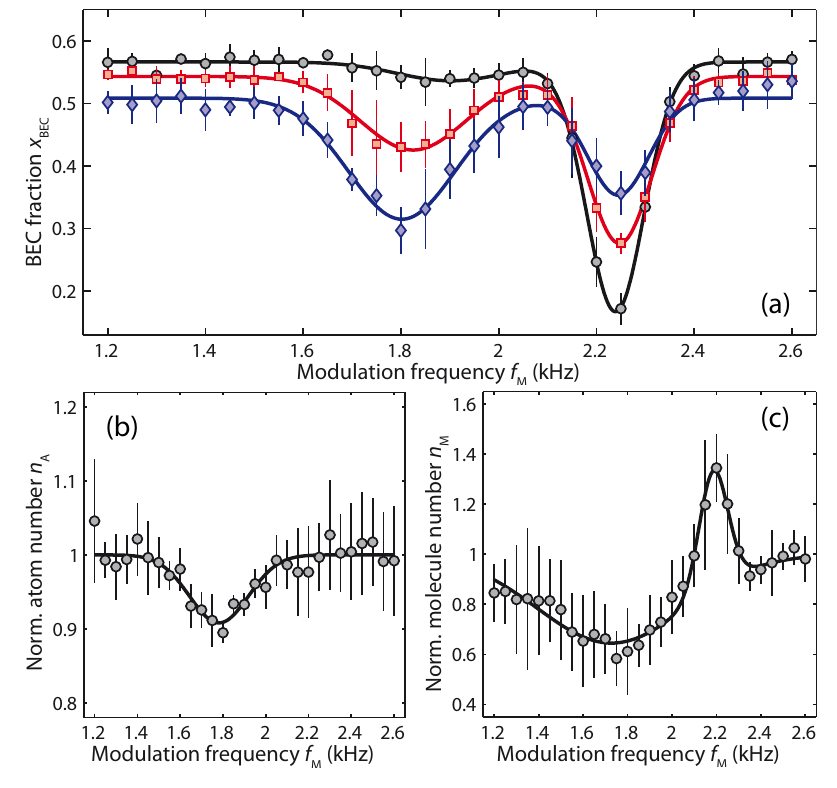}
\caption{(color online) a) Detailed structure of the resonance near $U$ for different values of the initial atom density at $a_{\rm S}\,{=}\,427\,a_0$ (low density, no double occupancy (circles), intermediate density, some double occupancy (squares), high density, high double occupancy (diamonds), for details see text). Every data point is the statistical average of five measurements, the error bars are the standard deviation. b) Remaining atom number $n_{\rm A}$ after AM at the same interaction strength as in a), normalized to $n_{\rm A}$ without AM. The solid line is a gaussian fit. c) Remaining molecule number $n_{\rm M}$ after AM at the same interaction strength as in a), normalized to $n_{\rm M}$ without AM. The solid line is a double gaussian fit.\label{Fig3}}
\end{figure}

Fig.\,\ref{Fig2}(c) shows an excitation spectrum as we increase the effect of interactions ($a_{\rm S}\,{=}\,320\,a_0$). Evidently, the resonance corresponding to $U$ splits into two clearly visible but not yet fully resolved components. The resonance at $U/2$ develops a shoulder on the high-frequency side. The splitting becomes more pronounced for even higher values of $a_{\rm S}$. A detailed excitation spectrum around $U$ is shown in Fig.\,\ref{Fig3}(a) for three different values for the initial density ($a_{\rm S}\,{=}\,427\,a_0$). The two components are now well separated from each other. Their strength depends in opposite ways on the initial density: The resonance at lower frequency (R$_1$) decreases in strength when the initial density is reduced and nearly disappears at low densities, whereas the resonance at higher frequency (R$_2$) increases in strength. We interpret this behavior in the following way: R$_1$ is caused by excitations in the doubly occupied Mott shell, whereas for R$_2$ singly occupied sites are excited. We also detect an increased atom loss in conjunction with R$_1$ as shown in Fig.\,\ref{Fig3}(b). Evidently, this resonance corresponds to excitations of doubly into triply occupied sites, thereby leading to three-body atom loss \cite{Weber2003a}. In Fig.\,\ref{Fig3}(c) we plot the number of atoms at doubly occupied sites measured directly by associating them to molecules via a Feshbach resonance, removing the unpaired atoms, and detecting the remaining fraction of molecules $n_{\rm M}$ \cite{Danzl2009,Danzl2010}. We observe a decrease of $n_{\rm M}$ at R$_1$ and an increase at R$_2$, in agreement with our interpretation for the origin of the two resonances.

\begin{figure}
\includegraphics[width=8.5cm]{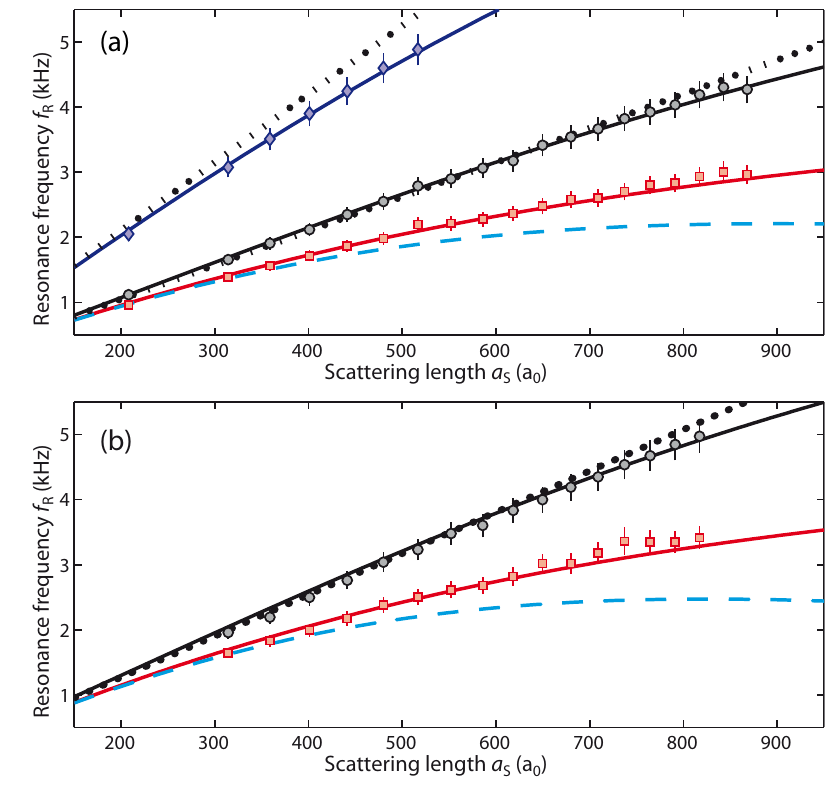}
\caption{(color online) a) Frequency of the upper (R$_2$, circles) and lower (R$_1$, squares) resonance around $U$ and of the resonance around $2\,U$ (diamonds) as a function of $a_{\rm S}$ for a lattice depth $V_0=20\,E_{\rm R}$. The dotted lines are the calculated $U(1)$ and $2\,U(1)$ values from the standard BH model. The solid lines are the result of the more elaborate calculations for $U(2)$ and $3\,U(3)-2\,U(2)$ (see text). The dashed line is the calculated $3\,U(3)-2\,U(2)$ using the renormalized perturbation theory \cite{Johnson2009}. b) Same as in a), but for $V_0=25\,E_{\rm R}$. \label{Fig4}}
\end{figure}

We map out the dependence of the resonances on the interaction strength by varying the magnetic field $B$ from the point at about $B\,{=}\,21\,$G, where the two resonances start to split, to $B\,{=}\,40\,$G, spanning the range from $a_{\rm S}\approx 200\,a_0$ to $a_{\rm S}\approx 900\,a_0$ \cite{Julienne2011}. As the loading of the lattice at high values for $a_{\rm S}$ leads to considerable heating and loss, we ramp into the MI phase at $a_{\rm S}\,{=}\,400\,a_0$, set $V_0\,{=}\,20\,E_{\rm R}$, and subsequently increase $a_{\rm S}$ to the desired value with a ramp speed of $5\,$G/ms. Fig.\,\ref{Fig4}(a) shows the frequencies for R$_1$ and R$_2$ and for the resonance at $2\,U$ as a function of $a_{\rm S}$. Computing the number dependent energies $U(2)$ and $U(3)$ is non-trivial because of the anharmonicity of the lattice potential and the need to regularize the $\delta$-function pseudopotential for the interactions. Values for $U(2)$ were calculated in Ref.~\cite{Buechler2010}, and for our parameters are also well approximated by rescaling the exact result for two atoms in a harmonic trap \cite{Busch1998} to correct for anharmonicity, using the ratio of the Born approximation results for the lowest oscillator levels in our lattice and the harmonic trap \cite{Mentink2009,Grishkevich2009}. We plot this in Fig.\,\ref{Fig4}(a) as the solid black line and see that it agrees well with our data for R$_2$. Note that an approach incorporating higher bands but not renormalizing the pseudopotential fails within our range of $a_{\rm S}$ values. $U(3)$ can be estimated using the renormalized perturbation theory of Ref.~\cite{Johnson2009}. This result is plotted as a dashed line in Fig.\,\ref{Fig4}(a), and agrees well for small values of $a_{\rm S}$. In order to reach larger values of $a_{\rm S}$ we would need to properly resum the perturbation expansion, which up to now is an open problem. Interestingly, we find that the function $3U(3)-2 U(2) \approx U(2)/\left(1+1.34 U(2)/(\hbar \omega)\right)$, with $\hbar \omega$ the band gap, equivalent at second order in $U(2)/(\hbar \omega)$ to the perturbation result, agrees well with our experimental data for R$_1$, as shown by the solid red line in Fig.\,\ref{Fig4}(a). The same measurements and calculations for a lattice depth of $25\,E_{\rm R}$ are shown in Fig.\,\ref{Fig4}(b), and we see again good agreement between our calculations and the experimental data. For comparison we plot also the interaction energies calculated with the standard BH model as dotted lines. Remarkably, this basic calculation fits our data for $U(2)$ even at intermediate interaction strengths, due to opposing corrections that arise from including higher bands and regularizing the pseudopotential.

We have investigated the SF-to-MI quantum phase transition and the MI phase over a large region of interaction strengths. For strong interactions, beyond single-band BH effects appear, leading to a splitting of the excitation resonances in the MI phase. Our precise measurements amount to a careful calibration of system parameters, including the scattering length over a wide range. The splitting of the excitation resonances can be used to manipulate the Mott shells independently, for example increasing the number of doubly occupied sites without loss due to excitation to triply occupied sites.

We are indebted to R. Grimm for generous support. We thank D. Boyanovsky, H. B\"uchler, P. Johnson, W. Niedenzu, and E. Tiesinga for fruitful discussions. We gratefully acknowledge funding by the Austrian Science Fund (FWF) within project I153-N16 and within the framework of the European Science Foundation (ESF) EuroQUASAR collective research project QuDeGPM.

\bibliographystyle{apsrev}

\begin{references}
\bibitem{Tunneling} Note that the admixture of higher bands can also change tunneling rates.
\bibitem{fit1} The critical depth $V_{\rm C}$ is identified as the intersection point of two linear functions that we add quadratically and fit to the data.
\end{references}

\end{document}